\documentclass[final, aps, prl, reprint, twocolumn, superscriptaddress]{revtex4-2}

\usepackage{amsmath}
\usepackage{amsfonts}
\usepackage{cleveref}
\usepackage{amssymb}
\usepackage{siunitx}
\DeclareSIUnit\Molar{\textsc{m}}
\DeclareSIUnit\kT{\textsc{k}_BT}
\DeclareSIUnit\vv{\!\%\, \text{vol}/\text{vol}}
\DeclareSIUnit\wv{\!\%\, \text{wt}/\text{vol}}
\DeclareSIUnit\kt{\textsc{k}_{\text{B}}\textsc{T}}
\DeclareSIUnit\st{\milli\newton\per\meter}
\usepackage{listofsymbols}
\usepackage{derivative}
\usepackage{array}
\newcolumntype{L}{>{$}l<{$}}
\newcolumntype{C}{>{$}c<{$}}

\usepackage{subfiles}
\usepackage{glossaries}
\glsdisablehyper 
\newglossaryentry{kbT}{
name={Energy Unit},
symbol={\ensuremath{\textsc{k}_{\text{B}}\textsc{T}}},
description={Boltzmann constant times Temperature}
}
\newglossaryentry{total_dna}{
	name={Total DNA},
	symbol={\ensuremath{N}},
	description={the total DNA on a droplet}
}
\newglossaryentry{complement_dna}{
	name={Compliment DNA},
	symbol={\ensuremath{N'}},
	description={the total DNA on the compliment droplet}
}
\newglossaryentry{patch_dna}{
	name={Patch DNA},
	symbol={\ensuremath{n}},
	description={the DNA in the patch}
}

\newglossaryentry{patch_thresh}{
	name={Patch DNA Threshold},
	symbol={\ensuremath{n_{C}}},
	description={the DNA in the patch}
}
\newglossaryentry{patch_area}{
	name={Patch Area},
	symbol={\ensuremath{A}},
	description={the area of the patch}
}
\newglossaryentry{area_limit}{
	name={Geometric Limit},
	symbol={\ensuremath{A_{\text{lim}}}},
	description={the geometric limit of the system}
}
\newglossaryentry{surface_area}{
	name={Surface Area},
	symbol={\ensuremath{A_0}},
	description={The surface area of the droplet}
}
\newglossaryentry{surface_tension}{
	name={Surface Tension},
	symbol={\ensuremath{\gamma}},
	description={The surface tension of the droplets}
}
\newglossaryentry{patch_conc}{
	name={Patch Concentration},
	symbol={\ensuremath{c_p}},
	description={Surface Density of DNA in the patch }
}
\newglossaryentry{droplet_conc}{
	name={Droplet Concentration},
	symbol={\ensuremath{c^{\phantom{'}}_0}},
	description={Surface Density of DNA on a droplet}
}
\newglossaryentry{partner_conc}{
	name={Droplet Concentration},
	symbol={\ensuremath{c^{\prime}_0}},
	description={Surface Density of DNA on the parner droplet}
}
\newglossaryentry{linker_area}{
	name={Linker Area},
	symbol={\ensuremath{a}},
	description={Area that characterizes the closeness needed to find a partner in the patch}
}
\newglossaryentry{ener_binding}{
	name={Binding Energy},
	symbol={\ensuremath{\mathcal{F}_b}},
	description={Energy of complimentary pairs of DNA binding}
}
\newglossaryentry{ener_spring}{
	name={Spring Energy},
	symbol={\ensuremath{\mathcal{F}_s}},
	description={Energy penalty of stretching the complex spring}
}
\newglossaryentry{ener_deformation}{
	name={Deformation Energy},
	symbol={\ensuremath{\mathcal{F}_{\gamma}}},
	description={Energy penalty of deforming the droplet}
}
\newglossaryentry{ener_interaction}{
	name={Onsauger Interaction Energy (from PNAS)},
	symbol={\ensuremath{\mathcal{F}_{r}}},
	description={Energy of Onsauger interaction binders}
}
\newglossaryentry{ener_total}{
name={Total Energy},
symbol={\ensuremath{\mathcal{F}}},
description={Total energy functional of the dimer system}
}
\newglossaryentry{config_micro}{
	name={Configurational Microstate},
	symbol={\ensuremath{\Omega_C}},
	description={Configurational Microstate of the patch}
}
\newglossaryentry{const_bind}{
	name={Energy Constant of Binding},
	symbol={\ensuremath{\epsilon}},
	description={Energy constant of binding}
}
\newglossaryentry{const_spring}{
	name={Energy Constant of Spring},
	symbol={\ensuremath{s_0}},
	description={Energy constant of Spring}
}
\newglossaryentry{bind_eff}{
	name={Energy Constant of Spring},
	symbol={\ensuremath{\epsilon_{\text{eff}}}},
	description={Energy constant of Spring}
}
\newglossaryentry{ener_tot_def}{
	name={Total Deformed Regime Energy},
	symbol={\ensuremath{\mathcal{F}_{\textbf{D}}}},
	description={Total free energy functional of the deformed regime}
}
\newglossaryentry{ener_tot_undef}{
	name={Total Undeformed Regieme Energy},
	symbol={\ensuremath{\mathcal{F}_{\textbf{U}}}},
	description={Total free energy functional of the undeformed regime}
}
 
\usepackage{graphicx}
\graphicspath{{./img/}}

\usepackage{xcolor}

\begin{document}

\author{Nicolas Judd}
\affiliation{Center for Soft Matter Research, New York University, New York, NY 10003}
\author{Angus McMullen}
\affiliation{Center for Soft Matter Research, New York University, New York, NY 10003}
\author{Sascha Hilgenfeldt}
\affiliation{Mechanical Science and Engineering, University of Illinois at Urbana-Champaign, Urbana, IL 61801}
\author{Jasna Brujic}
\affiliation{Center for Soft Matter Research, New York University, New York, NY 10003}

\title{A thermodynamic approach to adhesion and deformation of DNA-bound droplets}

\begin{abstract}
	Here we derive and experimentally test a free energy functional that captures the adhesion of DNA-coated emulsion droplets. Generalizing previous approaches, the theory combines important energetic and entropic effects of microscopic DNA mechanics and droplet elasticity. It simultaneously predicts adhesion size, morphology, and binder concentration as a function of experimental control parameters. Notably, droplets transition from undeformed binding to flat droplet interfaces at a characteristic DNA coverage. These equilibrium predictions agree quantitatively with experiments on droplet-substrate and droplet-droplet binding, revealing a weak effective binding strength of $3.7\pm0.3\glssymbol{kbT}$ owing to entropic costs. Our results open the path to rich design strategies for making colloidal architectures.
\end{abstract}

\maketitle

Biological cells are a prime example of a system in which membrane deformation
and molecular adhesion dictate cellular self-assembly into large-scale
structures, such as tissues~\cite{foty_differential_2005}. Their shape
distribution and adhesion strength are crucial to
biological function~\cite{butler_cell_2009}. Analogously, droplets~\cite{hadorn_dna-mediated_2010},
vesicles~\cite{smith_progress_2009, parolini_volume_2015-1}, and
colloids~\cite{van_der_meulen_solid_2013, dreyfus_simple_2009,
	varilly_general_2012-1} coated with ligands are known to self-assemble into
programmable architectures, such as chains~\cite{mcmullen_dna_2021},
foldamers~\cite{mcmullen_self-assembly_2022},
clusters~\cite{kim_embryonic_2021, mitra_coarse-grained_2023-1}, soft
gels~\cite{janmey_negative_2007}, and biomimetic
tissues~\cite{nagendra_push-pull_2023}. These systems offer avenues for biomimicry in simplified model systems to help
understand mechanisms underlying biological processes~\cite{vian_situ_2023,
	pontani_biomimetic_2012-2, kim_embryonic_2021, casas-ferrer_design_2021}. Moreover, they open the path to tunability in materials with novel mechanical
and optical properties ~\cite{he_colloidal_2020-2,
	dillavou_demonstration_2022-1, hexner_effect_2020}.

On the molecular scale, the binder strength, flexibility, and specificity of
interactions have been shown to influence the structure of self-assembled
particulate networks~\cite{feng_specificity_2013-4, mitra_coarse-grained_2023-1}. On the scale of the
particles, increasing the concentration of binders at the interface leads to
more connected networks with a higher valence
~\cite{michele_developments_2013} and larger
deformations away from spherical~\cite{pontani_biomimetic_2012-2}.
Further, decreasing the stiffness of the particles or the surface tension
of emulsion droplets allows for the formation of cohesive packings that mimic
tissues~\cite{nagendra_push-pull_2023}.

To gain greater control over these systems, it is necessary to construct
a theoretical model that relates molecular binder properties
to the self-assembly of the constituent particles. The literature spans
from microscopic models of protein-mediated cell adhesion~\cite{bell_models_1978-2,
	bell_cell_1984, dembo_thermodynamics_1987, fenz_membrane_2017} and cellular
recognition~\cite{qi_synaptic_2001}, to DNA-binding of
vesicles~\cite{parolini_volume_2015-1}, and biomimetic droplets
\cite{pontani_biomimetic_2012-2, mcmullen_dna_2021}.
These models balance ligand binding strength with particle elasticity to reach mechanical equilibrium, which is distinct from adhering solid particles, where the ligands are immobile and the binding region is limited by the spherical geometry of the interfaces \cite{dreyfus_simple_2009, varilly_general_2012-1, biancaniello_colloidal_2005-1, wang_synthetic_2015}.
In the case of droplet-droplet binding, low concentrations of DNA result in adhesions with extended linkers between spherical interfaces \cite{mcmullen_dna_2021}. Adding more DNA binders to the droplet surfaces leads to progressively larger and denser adhesion patches, which can deform into flat interfaces if the binders overcome the surface tension cost.

\begin{figure}[ht] \centering
	\includegraphics[width=\columnwidth]{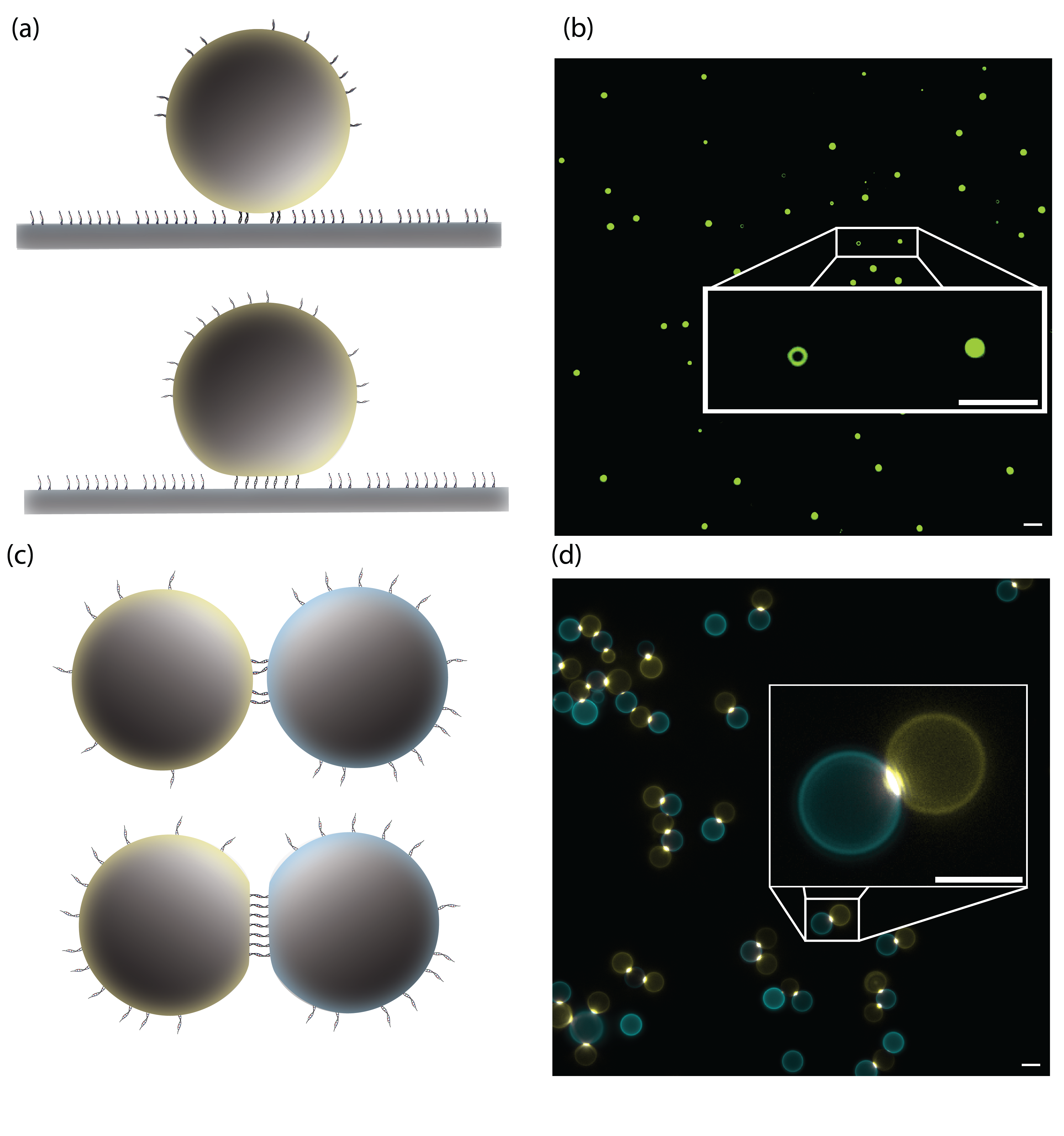}
	\caption{ Schematics of undeformed (top) and deformed (bottom) droplet adhesion via DNA linkers in the droplet-substrate (a) and droplet-droplet (c) experiments. Fluorescence microscopy images of droplet-substrate (b) and droplet-droplet (d) binding show a distribution of adhesion patches. Zooms show ring and disk adhesion morphologies on the substrate, while dimer droplets exhibit both perimeter and patch fluorescence. Scale bars are $\qty{5}{\um}$. }\label{fig: schematics}
\end{figure}

Here we combine experiments and theory to develop a thermodynamic model for the deformed droplet regime, as shown in the schematics of \cref{fig: schematics}. The cost of the energy of deformation, governed by surface tension $\glssymbol{surface_tension}$, balances with the configurational entropy gain of the binders to give the equilibrium patch DNA density and size as a function of the total numbers of DNA $\glssymbol{total_dna}$ and $\glssymbol{complement_dna}$ on the droplet surfaces. Comparing the free energies of the undeformed \cite{mcmullen_dna_2021} and deformed bound droplets allows us to identify the DNA coverage that favors the deformed state. By varying surfactant concentration, we show that emulsions with a lower surface tension require less DNA to deform into flat patches, consistent with model predictions. Once deformed, the area of adhesion $\glssymbol{patch_area}$ grows approximately with the square root of the number of DNAs $\glssymbol{patch_dna}$ inside the adhesion patch, in good agreement with theory. While the nominal DNA binding strength $\glssymbol{const_bind}$ plays an important role in determining the fraction of DNA recruited from the droplet surface into the patch, it does not significantly affect patch size. Thus, our statistical mechanics approach describes equilibrium droplet adhesion spanning spherical and deformed regimes, allowing one to tune macroscopic adhesion shape and strength from the bottom up.

The free energy functional of the deformed system consists of the energy gain of DNA hybridization, the energy costs of binder stretching, crowding, and surface deformation, and the configurational entropy changes due to binder recruitment from the surface into the adhesion patch,

\begin{equation}\label{eq: f_sum}
	\glssymbol{ener_tot_def} = \glssymbol{ener_binding} + \glssymbol{ener_spring} +
	\glssymbol{ener_interaction} +
	\glssymbol{ener_deformation} - \glssymbol{kbT} \ln{\glssymbol{config_micro}}.
\end{equation}

\noindent
At high droplet coverage and significant deformation, the theory assumes that the DNA linkers stand perpendicular to the bound
interfaces, implying that the steric repulsion between the strands $\glssymbol{ener_interaction}$ is
negligible. Additionally, the Debye length in our system is on the order of
$\qty{1}{\nm}$, such that electrostatic repulsion can be ignored.

The total hybridization energy $\glssymbol{ener_binding} = -
	\glssymbol{const_bind} \glssymbol{patch_dna}$ is reduced by a uniform molecular
stretching penalty $\glssymbol{ener_spring} =
	\glssymbol{const_spring}\glssymbol{patch_dna}$, where $\glssymbol{patch_dna}$
is the number of DNAs in a patch. The resulting effective binding energy is
balanced against the cost of droplet deformation, modeled as

\begin{equation}\label{eq: def_energy}
	\glssymbol{ener_deformation} = \nu\glssymbol{surface_tension}
	\frac{\glssymbol{patch_area}^2}{\glssymbol{surface_area}}
\end{equation}

\noindent
where $\nu$ is the number of deforming surfaces, $\glssymbol{surface_tension}$ is the surface tension and $\glssymbol{patch_area}$ and $\glssymbol{surface_area}$ are the areas of the deformed patch
and the undeformed droplet surface, respectively \cite{princen_rheology_1983}. Equation (\ref{eq: def_energy}) is accurate if $\glssymbol{patch_area} \ll \glssymbol{surface_area}$.
Finally, configurational entropy is modeled by counting the number of microstates
$\glssymbol{config_micro}$ that $\glssymbol{total_dna}$ and  $\glssymbol{complement_dna}$ DNA binders
on the two surfaces can assemble into

\begin{equation}\label{eq: config_microstate}
	\glssymbol{config_micro}\!=\! \binom{\glssymbol{total_dna}}{\glssymbol{patch_dna}}
	\binom{\glssymbol{complement_dna}}{\glssymbol{patch_dna}}
	\glssymbol{patch_dna}!
	\glssymbol{patch_area}^{\glssymbol{patch_dna}}\glssymbol{linker_area}^{\glssymbol{patch_dna}}
	(\glssymbol{surface_area}\!-\! \glssymbol{patch_area})^{\glssymbol{total_dna}\!-\!\glssymbol{patch_dna}}
	(\glssymbol{surface_area}\!-\! \glssymbol{patch_area})^{\glssymbol{complement_dna}\!-\!\glssymbol{patch_dna}}
\end{equation}

\noindent
where $\glssymbol{linker_area} = \frac{1}{4}
	\left(\glssymbol{droplet_conc}{}^{-\frac{1}{2}}
	+\glssymbol{partner_conc}{}^{-\frac{1}{2}}\right)^2 $ is the linker area that
characterizes the closeness needed by a DNA binder to find a partner in the
patch, for initial DNA concentrations $\glssymbol{droplet_conc}=\glssymbol{total_dna}/\glssymbol{surface_area}, \glssymbol{partner_conc}=\glssymbol{complement_dna}/\glssymbol{surface_area}$ \cite{SM}.

Minimizing the free energy with respect to $\glssymbol{patch_area}$ and $\glssymbol{patch_dna}$ respectively gives
\begin{equation}\label{eq: min_Z}
	\pdv{\glssymbol{ener_tot_def}}{\glssymbol{patch_area}}\!=\!0\!=
	2\nu\glssymbol{surface_tension}
	\frac{\glssymbol{patch_area}}{\glssymbol{surface_area}} -
	\frac{\glssymbol{patch_dna}}{\glssymbol{patch_area}} \glssymbol{kbT}  +
	\frac{(\glssymbol{total_dna} + \glssymbol{complement_dna} -
		2\glssymbol{patch_dna})}{\glssymbol{surface_area} - \glssymbol{patch_area}}
	\glssymbol{kbT}
\end{equation}
and
\begin{equation}\label{eq: min_n}
	\pdv{\glssymbol{ener_tot_def}}{\glssymbol{patch_dna}} =0=
	-\glssymbol{bind_eff} +
	\glssymbol{kbT}\ln{\left(
		\frac{\frac{\glssymbol{patch_dna}}{\glssymbol{patch_area}\glssymbol{surface_area}}}{1-\frac{\glssymbol{patch_dna}}{\glssymbol{total_dna}}}
		\frac{\left(\frac{\glssymbol{surface_area}-\glssymbol{patch_area}}{\sqrt{\overline{\glssymbol{total_dna}}}}\right)^2}{1-\frac{\glssymbol{patch_dna}}{\glssymbol{complement_dna}}}
		\right)},
\end{equation}

\noindent
with $\glssymbol{bind_eff} = \glssymbol{const_bind} - \glssymbol{const_spring}$ and the mean $\overline{\glssymbol{total_dna}} =
	\left(\left(\glssymbol{total_dna}^{\frac{1}{2}} +
		\glssymbol{complement_dna}{}^{\frac{1}{2}}\right)/2\right)^2$.
Simultaneously solving equations (\ref{eq: min_Z}) and (\ref{eq: min_n}) predicts $\glssymbol{patch_area}$ and $\glssymbol{patch_dna}$ as a function of experimental control parameters, $\glssymbol{total_dna}$, $\glssymbol{complement_dna}$, and $\glssymbol{surface_tension}$. Note that when
the density of DNA outside the patch is low, \cref{eq: min_Z} simplifies
to a square root law for the growth of $\glssymbol{patch_area}$ with
$\glssymbol{patch_dna}$,
\begin{equation} \label{eq: SN_model}
	\glssymbol{patch_area}  \approx \sqrt{\frac{\glssymbol{surface_area}\glssymbol{kbT}}{2\nu\glssymbol{surface_tension}} }
	\sqrt{\glssymbol{patch_dna}}.
\end{equation}
Interestingly, the patch area does not depend on the DNA binding energy in this limit.

\begin{figure}[ht]
	\includegraphics[width=\columnwidth]{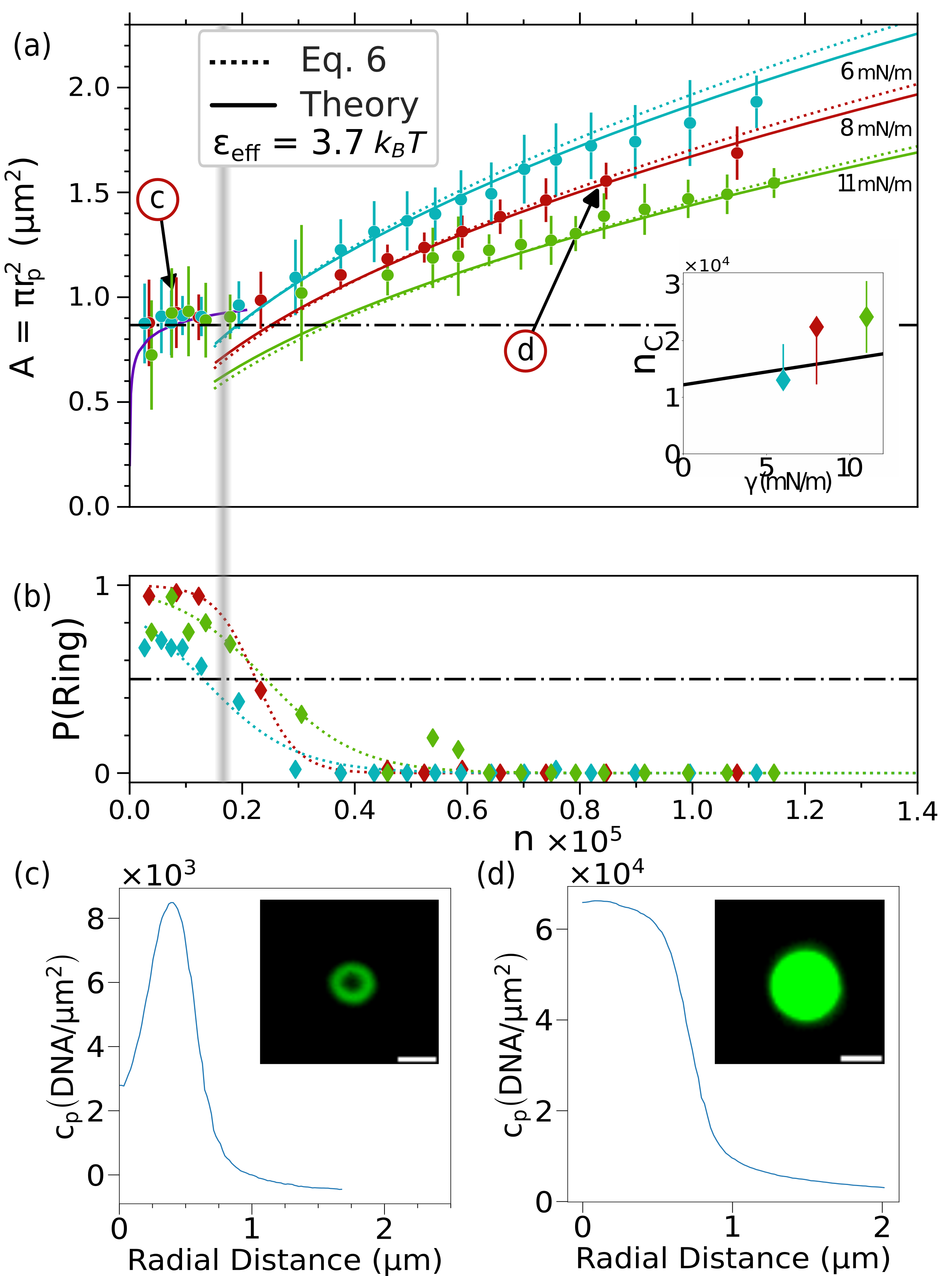}
	\caption{(a) Growth of droplet-substrate adhesion patch area $\glssymbol{patch_area}$ with $\glssymbol{patch_dna}$ exhibits a fast rise up to the undeformed droplet limit plateau (dot-dashed line), followed by a gradual increase in the deformed regime, in good agreement with theory and \cref{eq: SN_model}. Data is shown for 1, 2, and 3 mM TMN cosurfactant concentrations in green, red, and blue, respectively. The transition $\glssymbol{patch_thresh}$ is predicted to increase with surface tension (inset black line and grey shaded region).  (b) Ring probability P(Ring) sharply decreases with $\glssymbol{patch_dna}$, giving experimental estimates for $\glssymbol{patch_thresh}$ (inset in panel (a)). Panels (c) and (d) show the radial DNA profiles in ring and disk morphologies (indicated by arrows in panel (a)). Scale bars are $\qty{1}{\um}$.}
	\label{fig: drop-sub}
\end{figure}

\begin{figure*}[ht]
	\centering
	\includegraphics[width=179mm]{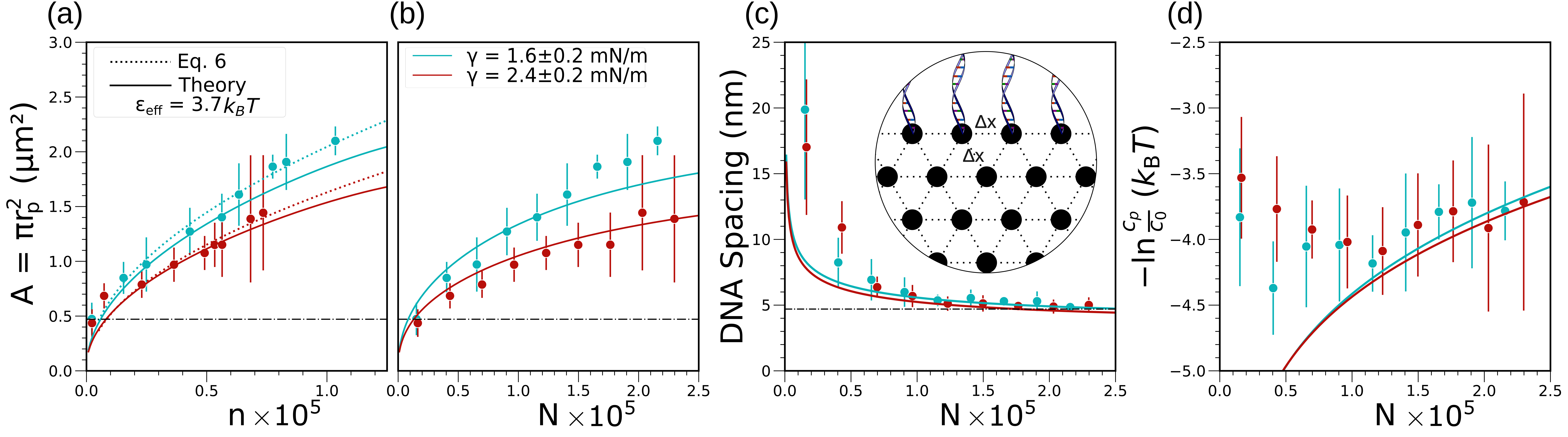}
	\caption{Droplet-droplet adhesion patch area $\glssymbol{patch_area}$ grows with the amount of DNA in the patch (a) and the total droplet DNA (b) for two emulsion batches. Experimental data are fit with the theory to give estimates for the surface tension and the effective binding strength. Fits with the simplified approach Eq.~\cref{eq: SN_model} are also shown for comparison. Note that the geometric limit (dot-dash line) is at the resolution of the microscope in this case. (c) DNA spacing in the adhesion patches (inferred from $n,A$ assuming a triangular lattice) approaches the close-packing limit of DNA (dashed line) with increasing DNA coverage ~\cite{koltover_dna_2000}. (d) The natural logarithm of the concentration ratio inside and outside the patch shows an increase at large $N$ due to crowding in both experiments and theory.     }\label{fig: drop-drop}
\end{figure*}
\twocolumngrid

To test the model, DNA-coated silane droplets stabilized with SDS surfactant are bound to a substrate surface coated with mobile complementary DNA of concentration $c_s$ (\cref{fig: schematics}(a)). In this case, the theory remains quantitatively valid when setting $\nu=1$ and $\glssymbol{complement_dna}=c_s\glssymbol{surface_area}$. High-intensity fluorescent adhesions are observed under the microscope (\cref{fig: schematics}(b)), resolving the patch size
$\glssymbol{patch_area}$ and radial distribution of DNA,
whose integral gives $\glssymbol{patch_dna}$. These data reveal the onset of droplet deformation by the change in morphology from ring-shaped to disk-shaped adhesions, as shown by the examples in the zoom in \cref{fig: schematics}(b). Given that the droplets are above the focal plane, it is not possible to measure their perimeter intensity, from which $\glssymbol{total_dna}$ could be inferred. Therefore, a second set of experiments is performed using droplet-droplet dimers ($\nu=2, $~\cref{fig: schematics}(c)) functionalized with complementary DNA with distinct fluorescent labels (\cref{fig: schematics}(d)). Here the intensities of both the patches and the droplet perimeters are simultaneously quantified, determining $\glssymbol{total_dna},\glssymbol{complement_dna}$ as well as $\glssymbol{patch_dna}$.

We use the droplet-substrate data to test the effect of surface tension on the transition from undeformed to deformed binding and subsequent patch growth. In \cref{fig: drop-sub}(a) we plot $\glssymbol{patch_area}$ versus $\glssymbol{patch_dna}$ for droplets with varying amounts of TMN co-surfactant. All data collapses onto a curve with a limiting patch area $\glssymbol{area_limit}= 0.9\mu$m$^2$ at low $\glssymbol{patch_dna}$, consistent with the geometric undeformed area $\glssymbol{area_limit}=2\pi R_0L$, where $R_0=3.0\pm0.3\mu$m is the droplet radius and $L=46$nm is the theoretical linker contour length, cf.\ \cite{mcmullen_dna_2021}. The behavior changes qualitatively for higher $\glssymbol{patch_dna}$, showing continuous growth above $\glssymbol{area_limit}$. This growth is well fit by  \cref{eq: SN_model}
using increasing surface tensions with decreasing TMN concentration, consistent with measured values~\cite{czajka_surfactants_2015}.
Patches with $\glssymbol{patch_area}>\glssymbol{area_limit}$ show a decreased likelihood of ring morphology
P(Ring)(\cref{fig: drop-sub}(b)), consistent with the transition from a spherical droplet to one with a flat interface. Sigmoidal fits show that $\text{P(Ring)}=0.5$ is crossed at higher
$\glssymbol{patch_dna}$ with increasing surface tension, as shown in the inset
in \cref{fig: drop-sub}(a). The error bars correspond to the gap in data around the transition point, which may be due to hysteretic effects.  \Cref{fig: drop-sub} Panels (c) and (d) show typical radial profiles of DNA before and after this transition, at points indicated on the graph in \cref{fig: drop-sub}(a).

Droplet-droplet binding data reveals a similar growth in the area versus patch DNA curve (\cref{fig: drop-drop}(a)) as the droplet-substrate case, in good agreement with the square root law in \cref{eq: SN_model} (dashed lines) for two emulsion batches. Lines of best fit correspond to surface tensions $\glssymbol{surface_tension}=2.8$ and $4.4\pm\qty{0.2}{\milli\newton\per\meter}$.
These droplets appear softer than those in the droplet-substrate case because of differences in surfactants~\cite{SM}. Due to a lower spatial resolution of patches, signatures of the morphology transition are not apparent. On the other hand, the total number of DNA at the interface $\glssymbol{total_dna}$ is readily measured, such that the full model solving Eqs. (\ref{eq: min_Z}) and (\ref{eq: min_n}) can be tested using both $\glssymbol{patch_dna}$ and $\glssymbol{total_dna}$. Fitting the data in \cref{fig: drop-drop}(a),(b) with the full model (solid lines) gives lower surface tensions $\glssymbol{surface_tension}=1.6$ and $2.4\pm\qty{0.2}{\milli\newton\per\meter}$ than \cref{eq: SN_model} (dashed lines), as well as an effective binding energy $\glssymbol{bind_eff}=3.7\glssymbol{kbT}\pm0.3\glssymbol{kbT}$. This discrepancy is because the full model additionally takes into account DNA recruitment and crowding effects inside the patch.
Indeed, the DNA spacing, obtained from the triangular lattice spacing per molecule $(2\glssymbol{patch_area}/\sqrt{3}\glssymbol{patch_dna})^{\frac{1}{2}}$ inside the patch, asymptotically reaches the crystalline limit of $\qty{4.7}{\nm}$~\cite{koltover_dna_2000} in \cref{fig: drop-drop}(c). This packing density limit fixes the maximum patch size possible with a given DNA binder.
Growing larger patches would require a decrease in surface tension or lateral attractive interactions~\cite{nagendra_push-pull_2023}.

Commonly, an effective binding energy per molecule is obtained from the logarithmic ratio of concentrations inside and outside the patch, shown in \cref{fig: drop-drop}(d). Both in experiments and theory, this ratio decreases at large  $\glssymbol{total_dna}$. This is because the further growth of a patch becomes less favorable due to increased crowding.
The obtained values are in the range of a few $\glssymbol{kbT}$, consistent with those reported for the same DNA binders in the undeformed regime in \cite{mcmullen_dna_2021} where droplet-droplet binding was shown to be reversible. While this ratio varies with $\glssymbol{total_dna}$, the quantity $\glssymbol{bind_eff}$ is a property of the individual DNA molecules as modeled in \cref{eq: f_sum}. Its fitted value of $3.7\glssymbol{kbT}$ is much lower than the expected $-\Delta G\approx 48\glssymbol{kbT}$ \cite{fornace_nupack_2022, zadeh_nupack_2011} for the DNA sequence used in this study
because it includes the conformational entropy loss of DNA molecules confined in the binding patch, as well as the stretch energy per molecule $\glssymbol{const_spring}$. For large $\glssymbol{total_dna}$, the logarithmic concentration ratio inferred from our theory with this value of $\glssymbol{bind_eff}$ is in good agreement with experiment (\cref{fig: drop-drop}(d)).

Given the fact that we use the same DNA in both sets of experiments, we can use the estimated value for $\glssymbol{bind_eff}$ in the full model for droplet-substrate binding. The solid lines in \cref{fig: drop-sub}(a) show that the model successfully predicts $\glssymbol{patch_area}$  for large $\glssymbol{patch_dna}$. In order to estimate the patch coverage where the droplet first deforms, we now compare the free energies $\glssymbol{ener_tot_def}$ of the present formalism with those of the undeformed droplet theory $\glssymbol{ener_tot_undef}$ \cite{mcmullen_dna_2021}. The latter differs from \cref{eq: f_sum} in two respects: the microstate count $\glssymbol{config_micro}$ is evaluated in the limit of small $\glssymbol{patch_dna}$, and the molecular contributions $\glssymbol{ener_spring}, \glssymbol{ener_interaction}$ vary with position in the patch. This is because the spacing $h$ between interfaces varies with the position for undeformed droplets, resulting in non-uniform concentration profiles (cf.\ \cref{fig: drop-sub}(c)).
The functional form of the nonlinear molecular spring energy $s$ and the exclusion interaction between molecules are kept the same in the deformed case. Note that we do not neglect $\glssymbol{ener_interaction}$ in either $\glssymbol{ener_tot_undef}$ or $\glssymbol{ener_tot_def}$ here, as the patch coverage $\glssymbol{patch_dna}$ at the undeformed/deformed transition is relatively small. The uniform equilibrium distance $h_0$ between droplet and substrate in the deformed case is obtained by minimizing $\glssymbol{ener_tot_def}$ with respect to $h$, from which the equilibrium spring energy per molecule $\glssymbol{const_spring}=s(h_0)$ is deduced.

\begin{figure}[ht]
	\includegraphics[width=\columnwidth]{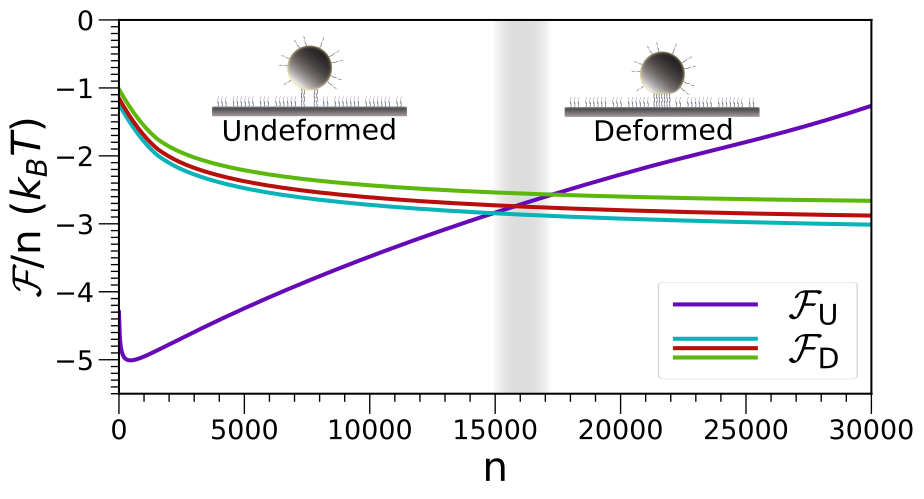}
	\caption{Free energy per molecule as a function of patch DNA $\glssymbol{patch_dna}$ in the undeformed $\glssymbol{ener_tot_undef}$ and the deformed $\glssymbol{ener_tot_def}$ cases, calculated using experimentally deduced values of $\glssymbol{bind_eff}$ and $\glssymbol{surface_tension}$ in the droplet-substrate case. The grey shaded region denotes the predicted transition where $\glssymbol{ener_tot_undef}/\glssymbol{patch_dna}$ = $\glssymbol{ener_tot_def}/\glssymbol{patch_dna}$, also shaded in \cref{fig: drop-sub}(a). $\glssymbol{ener_tot_def}$ colors correspond to droplet surface tensions shown in \cref{fig: drop-sub}.}
	\label{fig: ener_landscape}
\end{figure}

\Cref{fig: ener_landscape} plots the free energies per molecule $\glssymbol{ener_tot_undef}/\glssymbol{patch_dna}$ and $\glssymbol{ener_tot_def}/\glssymbol{patch_dna}$ for the three surface tensions of \cref{fig: drop-sub}(a). This identifies the $\glssymbol{surface_tension}$-dependent transition patch coverage $\glssymbol{patch_thresh}$ (gray shading, also shown in \cref{fig: drop-sub}(a),(b)). The functional dependence $\glssymbol{patch_thresh}(\glssymbol{surface_tension})$ predicted from theory is shown as the solid line in the inset of \cref{fig: drop-sub}(a), in good agreement with the empirical transition values obtained from experimental patch morphology data.

If we consider $\glssymbol{ener_total}/\glssymbol{patch_dna}$ to control the melting temperature of a patch, then \cref{fig: ener_landscape} predicts that melting temperature is maximized for sparsely covered undeformed patches. On the other hand, the total free energy of the patch, i.e. the binding strength, is significantly higher in the deformed regime with large $\glssymbol{patch_dna}$. The kinetics of the transition between the two states upon changes in $N$ remains an open question. However, the fact that  $\glssymbol{ener_total}/\glssymbol{patch_dna}$ is on the order of a few $\glssymbol{kbT}$ implies that the system is reconfigurable and will evolve towards the equilibria described here. Therefore, tuning the microscopic properties of the binders and the mechanical properties of the particles then allows for flexible control over the shape, size, and strength of adhesion of particles with mobile linkers. For instance, the spring energies could be replaced with Hookean springs or catch bonds~\cite{manibog_resolving_2014} in the case of protein-protein adhesion, or additional lateral interactions could be present, as in the case of DNA-condensation~\cite{teif_condensed_2011} or {\it cis}-bound cadherins~\cite{katsamba_linking_2009}. Alternatively, droplets could be exchanged for soft particles with Hertzian contact mechanics \cite{makse_granular_2004} or liposomes~\cite{liu_deformation_2006} with membrane bending elasticity. These changes to the free energy functional open the path to designing an even broader variety of adhesive particles. Extending the theory to higher coordination numbers will give rise to novel particulate networks (e.g.\ colloidal gels) with well-defined tunable architectures.

This work was supported by the NSF DMR grant No. 2105255 and the Swiss National Science Foundation through Grant No. 10000141. We thank J\'{e}r\^{o}me Bibette, Jean Baudry, and Frank Scheffold for insightful discussions.

\nocite{winkler_deformation_2003}
\nocite{mcmullen_freely_2018-1}
\bibliographystyle{apsrev4-2}
\bibliography{citations, SM}

\appendix
\section{Appendix}
\subsection{Droplet synthesis}

Following the procedure outlined in \cite{mcmullen_freely_2018-1}, the droplets
were prepared with an amount ranging from 1 to $\qty{20}{\vv}$ monomer and $\qty{20}{\vv}$ ammonia was dissolved in deinoized(DI) water. The monomer portion
was comprised of di\-eth\-oxy\-di\-methyl\-silane (Sigma
Aldrich) for droplet-substrate binding and of a mixture of dimeth\-oxy\-dimethyl\-silane (Sigma Aldrich)
and 3-Chloro\-propyl\-methyl\-dimethoxy\-silane (Gelest), in 4:1 proportion, for droplet-droplet binding. The additional monomer in the droplet-droplet case is to ensure that these droplets sink to the bottom of their flow cells for observation. The droplets were washed of ammonia and reaction by-products by centrifugation or dialysis with $\qty{5}{\milli\Molar}$ SDS.

\subsection{DNA sequences}

A: azide Cy5 A GCA TTA CTT TCC GTC CCG AGA GAC CTA
ACT GAC ACG CTT CCC ATC GCT A TT GTG AAC TCT TGT GAA CTC

A': azide Cy3 A GCA TTA CTT TCC GTC CCG AGA GAC CTA ACT GAC ACG CTT CCC ATC GCT A
GA GTT CAC AAG AGT TCA CAA

CS: TAG CGA TGG GAA GCG TGT CAG TTA GGT CTC TCG GGA CGG AAA GTA ATG CT azide

\subsection{Functionalizing droplets with DNA}

The DNA-functionalized droplets were prepared following the methods in
\cite{mcmullen_freely_2018-1, mcmullen_dna_2021}. Briefly, the azide-labeled DNA, purchased from Integrated DNA Technologies, and 1,2-distearoyl-sn-glycero-3-\-phosphoethanolamine-\-N-\-[dibenzo\-cyclooctyl (polyethylene glycol)-2000](ammonium salt)(DSPE-PEG-DBCO, Avanti Polar Lipids) are prepared as mobile binders to functionalize the droplets as described in \cite{mcmullen_freely_2018-1, mcmullen_dna_2021}.

The mobile binders are diluted to known concentrations with a dilute droplet
sample in $\qty{50}{\milli\Molar}$ NaCl TE buffer,
$(\qty{50}{\milli\Molar}\,\text{NaCl}$, $\qty{10}{\milli\Molar}\,\text{Tris}$
$\qty{1}{\milli\Molar}$ ethylenediaminetetraacetic acid (EDTA)), to
functionalize a known quantity of droplets. After incubating, the droplets are washed in
$\qty{50}{\uL}$ NaCl TE buffer with $0.1\%$ Triton-165 to remove loosely
attached DNA and then washed several times in $\qty{50}{\milli\Molar}$ NaCl TE
buffer to remove the Triton-165 solution before being stored in stored in $\qty{50}{\milli\Molar}$
NaCl TE buffer.

\subsection{Microscopy}

The droplets were observed using the 100x oil lens on a Nikon Ti-E epifluorescence microscope. The A DNA carries Cy5 fluorescent dye while the
A\textsuperscript{$\prime$} DNA carries Cy3 fluorescent dye, allowing for their
imaging in separate channels.

\subsection{Custom built flow cell}
To prepare the flow cell for the binding experiments, glass
slides and coverslips are first cleaned in a $\sim2\%$ solution of Helmanex
(Sigma Aldrich) then plasma cleaned in an
O\textsubscript{2} plasma bath for $\qty{30}{\min}$. After cleaning, the slides
and coverslips are hydrophobized with hexamethyldisilazane (Sigma Aldrich).
For droplet-substrate experiments, two coverslips are secured to a hydrophobized
slide with UV glue to produce a chamber. A third coverslip is place over the
chamber and secured with UV glue as well.
For droplet-droplet experiments, a piece of parafilm is placed on a
hydrophobized glass slide upon which chambers approximately $\qty{1}{\mm}$ wide are cut out. A coverslip is placed over the chambers and
secured by heating the slide and parafilm.

\subsection{Droplet-Substrate experiment}

The droplet-substrate data was collected following the procedure from
\cite{mcmullen_dna_2021}. Briefly, a custom-built flow chamber is loaded with the fluorescent lipid-DNA complex, which forms a mobile substrate on the surface of the chamber, while the remainder is washed out. The chamber is then filled with droplets in a buffer containing $\qty{20}{\milli\Molar}$ MgCL\textsubscript{2}, $\qty{0.1}{\wv}$
Brij-35 surfactant, and $\qty{5}{\milli\Molar}$ Tris. Patches are found using fluorescent and
brightfield images obtained from a 100x oil immersion lens using a Hough
transform. The average intensity of the patch is found by summing the intensities of the patch pixels and then dividing by the number of patch pixels. The radial profiles are found by drawing a ray from the center of the patch to a distance 3 times the patch radius around the patch in $\qty{15}{\degree}$ increments. These measures are averaged to arrive at the final radial
profiles.

The $\glssymbol{patch_dna}$ value and DNA concentration are estimated by
considering that droplets with the highest concentration of DNA considered, i.e.
those with similar coverage to the substrate, have all the DNA inside the
patches and none at the perimeter. Multiplying out the patch area gives a
$\glssymbol{patch_dna}/\text{Average Intensity}$ scaling to arrive at a value of
$\glssymbol{patch_dna}$ for the measured droplet-substrate patches.

\subsection{Droplet-Droplet experiment}
The droplet-droplet data was collected following the procedure from
\cite{mcmullen_dna_2021}. Functionalized droplets of the A and A' species
are both brought into a custom built flow cell. The species A droplets
varied in DNA concentration while the A' species of droplets were prepared to
have the same amount of DNA in all conditions. The droplets were immersed in a buffer containing
$\qty{20}{\milli\Molar}$ MgCl\textsubscript{2} and $\qty{0.5}{\wv}$ F38 pluronic
surfactant and $\qty{5}{\milli\Molar}$ Tris at pH 8. Microscopic images are
analyzed using custom MATLAB software. First droplets are located within the
image using a circular Hough transform on the brightfield image augmented
through histogram stretching and Gaussian filtering to reduce noise. After
obtaining the positions of the droplets, the fluorescent images of the A species
DNA are used to measure the distribution of DNA around the perimeter of the
droplets by finding the maximum intensity pixels in a radial scan around the center of the droplet. A patch on a droplet is then found by differentiating the pixel intensity
along the droplet perimeter to locate sharp changes in intensity, which in turn identify the edges of the patch. The patch size is extracted from the length of the
patch region under the implicit assumption that the patch is symmetrical.

The $\glssymbol{total_dna}$ and $\glssymbol{complement_dna}=200,000$ estimate is made by
integrating the total intensity on a droplet, both inside and outside the
patch, for each droplet, and interpolating between the average of the total
intensity of the droplets at a given bulk DNA concentration and the average number of DNA per droplet calculated for that concentration, under the assumption that all DNA was adsorbed onto the droplet. This assumption is consistent with the fluorometry measures from
\cite{mcmullen_dna_2021}.

\subsection{Energy functional contributions from molecular mechanics}
The DNA binders used in the present study are identical to those described in \cite{mcmullen_dna_2021}, i.e., a dsDNA oligomer connected to two PEG molecules anchored in the surfaces of droplets or substrate. Each bound linker thus represents a series of springs whose total energy $s(h)$ depends on the distance $h$ between the surfaces bound together. The DNA spring nonlinearity is taken from \cite{winkler_deformation_2003}; see \cite{mcmullen_dna_2021} for details. The interaction energy per molecule is likewise given by a function $g(h)$, evaluating the Onsager excluded-volume energy of two DNA molecules  (as fixed-length rods) dependent on $h$. For the undeformed droplets of \cite{mcmullen_dna_2021} this function decreases linearly with $h$. For deformed droplets with flat patches, $h$ is universal for all binder molecules and the two contributions dependent on $h$ are
\begin{equation}
	\glssymbol{ener_spring} + \glssymbol{ener_interaction}=  \glssymbol{patch_dna} s(h) \glssymbol{kbT} + \frac{\glssymbol{patch_dna}^2}{c_m\glssymbol{patch_area}}g(h) \glssymbol{kbT}\,,
	\label{eq:Fmol}
\end{equation}
for a patch of size $\glssymbol{patch_area}$ containing $\glssymbol{patch_dna}$ binders, and $c_m=1/(Ld)$ is a normalization concentration given by the length $L$ and width $d$ of the binders. No other free energy contribution depends explicitly on $h$, so that the mimimization of (\ref{eq:Fmol}) with respect to $h$ yields an equilibrium surface-to-surface distance $h_0$ and corresponding values $s_0=s(h_0)$ and $g_0=g(h_0)$. Together with $\glssymbol{ener_binding}$ and $\glssymbol{ener_deformation}$ described in the main text, these are the free energy contributions from patch properties only.

\subsection{Entropy terms for molecules in and outside the binding patch}
Entropic cost of a configuration of molecules depends on the $\glssymbol{patch_dna}$ molecules in the patch as well as the $\glssymbol{total_dna}-\glssymbol{patch_dna}$ and $\glssymbol{complement_dna}-\glssymbol{patch_dna}$ molecules outside the droplets (in the case of droplet-droplet binding), or the concentration $c_s$ of molecules on the substrate (in the droplet-substrate case). Exact microstate counts for arbitrary configurations are elaborate, but can be simplified using the following assumptions appropriate for moderate to large coverage $\glssymbol{total_dna},\glssymbol{complement_dna}$ of droplets (resulting in relatively large patch populations $\glssymbol{patch_dna}$), and contrasting with the limit of small $\glssymbol{total_dna}$ (and $\glssymbol{patch_dna}$) employed in \cite{mcmullen_dna_2021}:
(i) The crowded patches do not leave much room for unbound linkers, so we assume all $\glssymbol{patch_dna}$ patch linkers in the patch area $\glssymbol{patch_area}$ to be bound, leaving $\glssymbol{total_dna}-\glssymbol{patch_dna}$ unbound linkers outside, on an area $\glssymbol{surface_area}-\glssymbol{patch_area}$; (ii) Onsager interaction of unbound linkers is negligible; (iii) the area $a$ that a single linker in the patch has to be confined to in order to bind with a linker on the other surface is not given by a constant of molecular geometry $a_{mol}$, as it was in \cite{mcmullen_dna_2021}. This is because the larger coverage densities in the present work translate to areas per molecule $<a_{mol}$, providing a stronger constraint.

Taking the initial densities (upon the beginning of patch formation) to be $\glssymbol{droplet_conc}=\glssymbol{total_dna}/\glssymbol{surface_area}$ and $\glssymbol{partner_conc}=\glssymbol{complement_dna}/\glssymbol{surface_area}$ for the two droplets, the limit explored here is that of $\glssymbol{droplet_conc},\glssymbol{partner_conc},c_s>c_{mol}\equiv 1/a_{mol}$. For the types of linkers and droplet sizes used, this translates to $\glssymbol{total_dna},\glssymbol{complement_dna},N_S\gtrsim 17,000$, comfortably fulfilled over the range of interest of deformed droplets in the present work. In this limit, the areas covered per molecule then have characteristic linear extent $\glssymbol{droplet_conc}{}^{-1/2}$, $(\glssymbol{partner_conc})^{-1/2}$. Overlap, and thus binding, occurs when an area given by the mean linear extent is covered by the linker, i.e., an effective area of
\begin{equation}
	\glssymbol{linker_area} = \frac{1}{4}
	\left(\glssymbol{droplet_conc}{}^{-\frac{1}{2}}
	+\glssymbol{partner_conc}{}^{-\frac{1}{2}}\right)^2\,,\end{equation}
as given in the main text. Equation (3) of the main text is then the microstate count of $\glssymbol{patch_dna}$ molecules binding into a patch of size $\glssymbol{patch_area}$ (and confined to $a$ upon binding), with $\glssymbol{total_dna}-\glssymbol{patch_dna}$ and $\glssymbol{complement_dna}-\glssymbol{patch_dna}$ distributed outside the patch.

An analogous computation for droplet-substrate binding results in equivalent formulae (up to an irrelevant constant contribution to energy), if the coverage of the partner droplet is taken to be $\glssymbol{complement_dna} = c_s \glssymbol{surface_area}$, and $\glssymbol{partner_conc}$ is replaced by $c_s$.

\subsection{Transition from undeformed to deformed at moderate $\glssymbol{total_dna}$}
Although we neglect $\glssymbol{ener_interaction}$ for simplicity for large $\glssymbol{total_dna},\glssymbol{complement_dna}$, for accurate comparison of the undeformed and deformed energy functionals we do take the interaction energies into account. The minimization of (\ref{eq:Fmol}) yields a value of $h_0$ very insensitive to $\glssymbol{patch_dna}$ (or the control parameters), resulting in a spring energy $\glssymbol{const_spring}\approx 2.6k_BT$. In comparing $\glssymbol{ener_tot_def}$ with $\glssymbol{ener_tot_undef}$, the former (from the present formalism) represents a limit of large $\glssymbol{total_dna},\glssymbol{complement_dna}$, while the latter (taken directly from \cite{mcmullen_dna_2021}) represents a limit of $\glssymbol{patch_dna}\ll \glssymbol{total_dna},\glssymbol{complement_dna}$. As we find that $\glssymbol{patch_dna}$ grows consistently with the droplet coverages, the two approaches are expected to be comparable for intermediate values. As an indicator for when the deformed state becomes preferable, we compute $\glssymbol{ener_tot_undef}/\glssymbol{patch_dna}$ and $\glssymbol{ener_tot_def}/\glssymbol{patch_dna}$, the energy changes upon adding another binder to the patch. While these stay negative over the range explored (the patch population $\glssymbol{patch_dna}$ grows as the droplet population(s) increase), there is a critical value $\glssymbol{patch_thresh}$ beyond which growing a deformed patch is favorable over growth of the undeformed patch (see Figs.~2a and 4 of the main text).

\subsection{Equilibrium states for large $\glssymbol{total_dna}$}
In contrast to \cite{mcmullen_dna_2021}, we are not simply comparing free energies at a fixed patch coverage $\glssymbol{patch_dna}$ in the present work. Instead, we find equilibrium values of $\glssymbol{patch_dna}$ using $\glssymbol{total_dna},\glssymbol{complement_dna},\glssymbol{const_bind},\glssymbol{surface_tension}$ (for the droplet-droplet case) or $\glssymbol{total_dna},c_s,\glssymbol{const_bind},\glssymbol{surface_tension}$ (for the droplet-substrate case) as the experimentally relevant control parameters. Thus, the theory curves of plots like Fig.~2(a) of the main text are parametric plots: For a given substrate coverage $c_s$ and binder species, $\glssymbol{total_dna}$ is varied and the equilibrium formulae (4) and (5) of the main text are used to quantify both $\glssymbol{patch_dna}$ and $\glssymbol{patch_area}$.

\subsection{Full model fitting procedure}

The full model was fit using data where $\glssymbol{total_dna} > 50,000$ to
ensure that only data from the deformed regime was considered. The data was
binned so that a measure of
uncertainty for the values could be considered across the range of N values in
the deformed regime. The full model was fit by minimizing the logarithm of the sum
of the squared quotients of all residuals over a bin's standard deviation of
observed values of $\glssymbol{patch_dna}$ and $\glssymbol{patch_area}$,
$\left(\frac{\glssymbol{patch_dna}-\hat{\glssymbol{patch_dna}}}{\sigma_{\glssymbol{patch_dna}}}\right)^2+\left(\frac{\glssymbol{patch_area}-\hat{\glssymbol{patch_area}}}{\sigma_{\glssymbol{patch_area}}}\right)^2$,
where $\hat{n}, \hat{A}$ are the model predicted values and $\sigma_{\glssymbol{patch_dna}},
	\sigma_{\glssymbol{patch_area}}$ are the standard deviations of
$\glssymbol{patch_dna}$ and $\glssymbol{patch_area}$ of the bin.


\end{document}